\documentclass[conference]{IEEEtran}
\IEEEoverridecommandlockouts
\usepackage[T1]{fontenc}
\usepackage{cite}
\usepackage{amsmath,amssymb,amsfonts}
\usepackage{algorithmic}
\usepackage{balance}
\usepackage{graphicx}
\usepackage{textcomp}
\usepackage{listings}
\usepackage{url}
\usepackage{todonotes}
\usepackage{hyperref}
\usepackage[super]{nth}

\usepackage{booktabs}
\usepackage{csquotes}
\usepackage{microtype}

\usepackage{color}
\usepackage{xcolor}

\definecolor{ckeywordcolor}{RGB}{127,0,85}
\definecolor{cstringcolor}{RGB}{42,0,255}
\definecolor{ccommentcolor}{RGB}{63,127,95}

\def\BibTeX{{\rm B\kern-.05em{\sc i\kern-.025em b}\kern-.08em
    T\kern-.1667em\lower.7ex\hbox{E}\kern-.125emX}}

\newcommand{\ovgu}{Otto-von-Guericke University}

\usepackage{xspace}
\newcommand{\dataset}{AndroidCompass\xspace}

\begin{document}

\title{\dataset{}: A Dataset of Android Compatibility Checks in Code Repositories}

\author{
\IEEEauthorblockN{Sebastian Nielebock, Paul Blockhaus, Jacob Krüger, Frank Ortmeier}
\IEEEauthorblockA{\ovgu{} Magdeburg, Germany\\
\{sebastian.nielebock | paul.blockhaus | jacob.krueger | frank.ortmeier\}@ovgu.de}
}
	

\maketitle

\begin{abstract}
	\looseness=-1
Many developers and organizations implement apps for Android, the most widely used operating system for mobile devices.
Common problems developers face are the various hardware devices, customized Android variants, and frequent updates, forcing them to implement workarounds for the different versions and variants of Android APIs used in practice.
In this paper, we contribute the \emph{Android Compatibility checkS dataSet} (\dataset{}) that comprises changes to compatibility checks developers use to enforce workarounds for specific Android versions in their apps.
We extracted 80,324 changes to compatibility checks from 1,394 apps by analyzing the version histories of 2,399 projects from the F-Droid catalog.
With \dataset{}, we aim to provide data on when and how developers introduced or evolved workarounds to handle Android incompatibilities.
We hope that \dataset{} fosters research to deal with version incompatibilities, address potential design flaws, identify security concerns, and help derive solutions for other developers, among others---helping researchers to develop and evaluate novel techniques, and Android app as well as operating-system developers in engineering their software.
	
\end{abstract}

\section{Introduction}
\label{sec:introduction}

\looseness=-1
\noindent
Android is the prevailing operating system for mobile devices worldwide,\footnote{\url{https://gs.statcounter.com/os-market-share/mobile/worldwide}} with many developers working on Android itself, vendor-specific customizations, and mobile apps for users.
Since September 2008, Android has been released in 11 major releases with 30 different Android API levels (as of January 2021).
Consequently, most developers have to react regularly to version updates of Android, which, for instance, add support for new sensors and devices, address security concerns, or replace functionality.
While such updates promise improvements, not all vendors and users update their Android system, for example, because they first need to adapt their own customizations or because the hardware does not support the Android version anymore.
As a result, Android exists in different versions, each with API-specific functionalities that are not available in other versions.

Thus, Android apps face compatibility issues, for instance, because they rely on functionality that is deprecated in a new release.
Developers have to handle such incompatibilities to make their apps available for as many devices as possible---which, however, is challenging as recently shown again for COVID-19 tracing apps~\cite{Ahmed2020}.
To deal with incompatibilities, Android enables developers to configure their apps for specific API levels and provide alternative implementations.
The latter is typically achieved by checking the Android version of the device and adapting the control flow accordingly.

Recent research~\cite{Wei2016,Wei2020,Li2018,He2018,Xia2020} has studied the prevalence of and fixing strategies for such incompatibilities.
However, except for the work of Scalabrino et al.~\cite{Scalabrino2019,Scalabrino2020}, no research has been conducted on how Android incompatibilities are handled during the evolution of an app and Android itself.
Such evolution is particularly interesting, since it allows researchers to investigate \emph{when} and \emph{in what order} incompatibilities were handled with what workarounds.

In this paper, we describe \dataset{}, a dataset of historical changes of compatibility checks in the source code of Android apps.
\dataset{} comprises 80,324 individual single-line code changes of Android compatibility checks and their respective meta data, which we collected from 1,394 projects of the F-Droid catalog.\footnote{\url{https://f-droid.org/en/}\label{fn:fdroid}}
We hope to foster research on how developers introduce, change, and eventually fix compatibility checks, ideally leading to automatic support for suggesting fixing patterns to developers.
Our dataset and all artifacts related to this paper are publicly available.\footnote{\url{https://doi.org/10.5281/zenodo.4428340}\label{fn:repo}}
\section{Android Compatibility Issues}
\label{sec:android_compat}

\looseness=-1
\noindent
The variety of functionalities available for different devices as well as variants of the Android API causes compatibility issues in apps.
In a general sense, an \emph{incompatibility} (or compatibility issue) refers to a \enquote{state of not being able to exist or work with another person or thing because of basic differences.}\footnote{\url{https://dictionary.cambridge.org/us/dictionary/english/incompatibility}} 
In our case, an Android incompatibility refers to the use of an Android API element (e.g., a method) in an app while that element is not available (or not behaviorally equal) in the Android API level of the underlying device.
Note that Android versions, such as \textsc{Kitkat} or \textsc{Oreo}, may consist of multiple Android API levels, usually denoted by increasing integers for newer levels.

Scalabrino et al.~\cite{Scalabrino2019,Scalabrino2020} distinguish between \emph{forward} and \emph{backward} incompatibility from the perspective of an app.
Forward incompatibility occurs when the app is used with a newer version (i.e., higher API level) and an API element becomes incompatible (e.g., the element is removed in the newer API level).
Backward incompatibility occurs when the app is used with an older version (i.e., lower API level) and an API element becomes incompatible (e.g., it was not yet implemented in the older API level).
The Android framework usually avoids (with some rare exceptions) forward incompatibility by not deleting, but deprecating, obsolete API elements---which is denoted backward compatibility from the perspective of the Android framework.
This claim has been supported by recent research.
For example, He et al.~\cite{He2018} have been able to execute 4,041 out of 4,697 apps on newer Android versions without any modifications.

\begin{lstlisting}[float=t,caption={An in-code Android compatibility check.},label={lst:version-check}]
if(Build.VERSION.SDK_INT < Build.VERSION_CODES.KITKAT){
	//call API before Kitkat
} else {
	//call API of Kitkat
}
\end{lstlisting}

\looseness=-1
App developers can handle incompatibilities by configuring the minimally and maximally allowed Android API levels in the \texttt{AndroidManifest.xml}.
However, configuring the maximal level is neither enforced nor recommended, since the Android framework usually avoids forward incompatibilities.\footnote{\url{https://developer.android.com/guide/topics/manifest/uses-sdk-element}}
By defining the range of allowed API levels, an app cannot be installed on devices using an API level outside of this range.
Still, since this range may be too coarse-grained, app developers frequently check the actual Android version in the code itself~\cite{He2018,Scalabrino2019,Scalabrino2020}. 
For example, the code in \autoref{lst:version-check} checks whether the Android API level (i.e., \texttt{Build.VERSION.SDK\_INT}) is below the predefined constant for \textsc{Kitkat} (i.e., \texttt{Build.VERSION\_CODES.KITKAT}) and changes the control flow accordingly. 
Note that Xia et al.~\cite{Xia2020} have found that only $\approx$38\,\% of such checks provide an actual alternative functionality, while most checks only disable functionality.
To avoid backward incompatibilities, Android provides newer functionalities with a support library since API level 26,\footnote{\url{https://developer.android.com/topic/libraries/support-library}} and since API level 28 this library is part of the Android \textsc{Jetpack} libraries denoted as \textsc{AndroidX}.\footnote{\url{https://developer.android.com/jetpack/androidx}}
Unfortunately, this support library only provides a subset of newer API functionalities.

\looseness=-1
Recent research focused on the prevalence of Android incompatibilities and tool support to detect these automatically, for instance, with FicFinder~\cite{Wei2016,Wei2020}, CiD~\cite{Li2018}, IctAPIFinder~\cite{He2018}, ACRyL~\cite{Scalabrino2019,Scalabrino2020}, and RAPID~\cite{Xia2020}.
Essentially, all these tools identify the used Android API elements (typically method calls) and map them to the API levels in which they are available. 
Afterwards, they check whether these API levels are within those allowed for the app (i.e., in the \texttt{AndroidManifest.xml}) and whether the usage is protected with conditional statements (e.g., \lstinline[basicstyle=\normalsize\ttfamily]|if|). 
Since these tools require control-flow information, they analyze the byte code of the app (i.e., the \texttt{*.apk}).
The results indicate that Android incompatibilities are prevalent with $\approx$25\,\% to $\approx$83\,\% of the analyzed apps having at least one compatibility issue~\cite{He2018,Scalabrino2019,Scalabrino2020,Xia2020}. 
Root causes for these incompatibilities are device-specific reasons (e.g., different hardware, customized operating system) as well as Android-specific reasons (e.g., API evolution with insufficient support, errors in the Android API)~\cite{Wei2016, He2018, Wei2020,Scalabrino2020}.
Mostly, the identified incompatibilities caused functional problems with the app deviating from its intended behavior, but also problems with the performance, security, or user experience~\cite{Wei2016,Scalabrino2020,Wei2020}.

\looseness=-1
The datasets used for those analyses mostly refer only to the latest version of the app at the time of the analysis. 
While this is sufficient to elaborate on the prevalence of incompatibilities, it does not help to shed light on the learning curve and evolution of compatibility handling in Android. 
For example, it remains unclear how long it takes developers to identify and fix incompatibilities with new Android versions and how frequently version checks are changed until they are in a robust state or replaced. 
The only exception is the work of Scalabrino et al.~\cite{Scalabrino2019,Scalabrino2020} who considered 19,291 snapshots of 1,170 apps. 
While having a comparable size to \dataset, using the data of Scalabrino et al. requires parsing and compiling of source code. 
This is not necessary for \dataset, since it requires textual filtering only. 
Moreover, \dataset{} has more fine-grained timestamps based on commit information, while Scalabrino et al. used pre-sampled timestamps that are far more fragmented. 
Finally, \dataset{} is easier to access and transfer, since it is a simple csv file.
\section{\dataset{}}
\label{sec:dataset}

\noindent
Nest, we describe our dataset and its construction process.

\subsection{Dataset Construction}

\looseness=-1
\noindent
Initially, we conducted a preliminary analysis of Android-spe{\-}cif{\-}ic API misuses, which, except for the web crawling step, was independent of the construction of \dataset{}. 
However, we derived our validation data from that analysis. 
In this context, we denote an API misuse as a deviant use of an API from the one that was intended by the developers, and that eventually leads to negative behavior of the software (e.g., a software crash or a performance issue)~\cite{Amann2019,Nielebock2020,Nielebock2020a}.  

At first, we manually analyzed commits that fixed API misuses in FOSS Android apps, which we collected from F-Droid.\textsuperscript{\ref{fn:fdroid}}
For this purpose, we implemented a web crawler using scrapy\footnote{\url{https://scrapy.org/}} to identify all URLs directing to a Git repository on GitHub, resulting in 2,399 initial repositories. 
We restricted our analysis to Git and GitHub, since these are the prevalent version-control and repository-hosting systems, respectively~\cite{Kalliamvakou2014}. 

To identify commits that involved a fix, we used PyDriller~\cite{Spadini2018} and extracted from the repositories all commits with a message containing the keywords \enquote{fix,} \enquote{issue,} or \enquote{bug.} These keywords were inspired by previous work~\cite{Sliwerski2005}.
We reduced the number of commits to make them manually assessable. Particularly, we ignored repositories that excessively used these keywords in their messages (i.e., $>$6\,\% of all messages for repositories with $>$1,000 commits, or $>$10\,\% otherwise).
Then, we extracted for each commit all lines with third-party API changes using our previous extraction mechanism~\cite{Nielebock2020a}.
Also, we considered only small commits that changed at most ten methods and that changed imports prefixed with \texttt{android.*}. 
Finally, we transformed commit messages with standard natural-language processing techniques and filtered whether they contain \enquote{android,} \enquote{api level,} or any of the Android version names.\footnote{\url{https://en.wikipedia.org/wiki/Android_version_history}} 
We derived these steps in an exploratory manner to construct a small, relevant, and manually assessable validation dataset (of limited generalizability). 

Then, the first three authors assessed the remaining 522 commits from 278 repositories independently to identify fixing commits of Android API misuses.
We determined that 132 of them represented API-misuse fixes.
Aligning to previous studies, we found that compatibility issues were prevalent among these misuses.
As a result, we marked each commit considering whether they added or changed a compatibility check (cf. \autoref{lst:version-check}).
We found that 67 of the 132 commit changes involved code for handling compatibility issues.

\begin{lstlisting}[float=t,caption={Our regular expression to identify Android compatibility checks.},label={lst:regex}]
^(([^\*/]*VERSION.SDK_INT[ ]+(<|<=|==|>=|>).*) |
([^\*/]*(<|<=|==|>=|>)[@~@]+[a-zA-Z0-9\.]*VERSION.SDK_INT.*))
\end{lstlisting}

\begin{figure*}
\centering
\includegraphics[width=.9\textwidth, trim=300 115 190 0, clip]{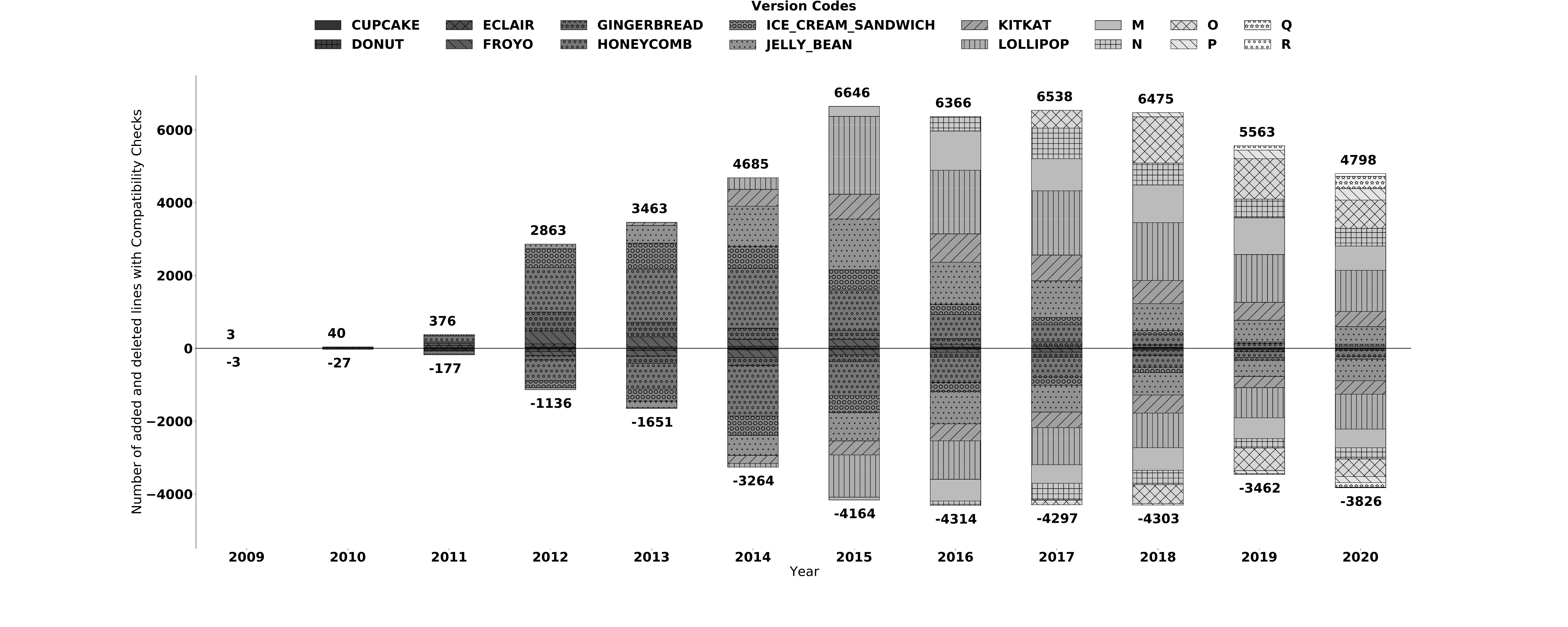}
\vspace*{-2ex}
\caption{A preliminary analysis of the historical development of added/deleted compatibility checks from 2009 to 2020.}
\label{fig:changed-version}
\vspace*{-2ex}
\end{figure*}

\looseness=-1
Based on our manual assessment, we derived the regular expression we show in \autoref{lst:regex} to automatically detect compatibility checks against the \texttt{VERSION.SDK\_INT}. 
This expression is part of a python script within our replication package.\textsuperscript{\ref{fn:repo}} 
Again, we used PyDriller to identify changed lines together with a set of meta-information (cf. \autoref{tb:field-desc}) for each line of code matching the regular expression.

\looseness=-1
We intended \dataset{} to provide information regarding which Android version was checked.
For this purpose, we extracted the version to which the \texttt{VERSION.SDK\_INT} constant is compared to and a normalized comparison type (i.e., \texttt{<|<=|==|>=|>} with the actual API level on the left).
To refine our dataset, we normalized the version codes, since developers can represent those, for instance, as numbers (e.g., 19), constants (e.g., \texttt{Build.VERSION\_CODES.KITKAT}), or results of a method call (e.g., \texttt{getKitkatCode()}).
We transformed numbers to version constants based on the Android documentation\footnote{\url{https://developer.android.com/reference/android/os/Build.VERSION_CODES}} (e.g., 19 refers to \texttt{Build.VERSION\_CODES.KITKAT}).
Similarly, we normalized varying version constants (e.g., \texttt{android.os.Build.VERSION\_CODES.KITKAT}) into that same format.
We marked versions as \texttt{NONE\_DETECTED} if we could not reliably ensure the correct version (e.g., in case of user-defined constants/methods or multiple version checks in a single line).
However, such corner cases occur sparsely within \dataset{} ($\approx$2.3\,\%).

We validated our analysis script based on the 132 commits we previously identified to comprise fixes for API misuses.
At this point, we denoted a detection as false negative if we assessed a commit to contain a compatibility check, but our script did not.
Vice versa, we denoted a detection as false positive if we did not identify a compatibility check, but our script did.
At first, we achieved a precision of 98\,\% and a recall of $\approx$73.1\,\%.
While validating these results, we found that the single false positive was actually a mistake in our manual assessment, meaning that the precision was 100\,\%.
Regarding false negatives, in most cases, PyDriller did not find the respective commit.
In other cases, our regular expression did not match, for instance, if the \texttt{VERSION.SDK\_INT} constant was stored in a local variable or if exceptions of the API were handled in an inherited, customized class.
We found two more cases that revealed errors in our regular expressions, which we fixed accordingly. 
After re-validating, we achieved an improved recall of $\approx$76.1\,\% with the same precision of 100\,\%.

Finally, we analyzed whether the extracted data is reasonable by comparing the proportion of lines adding a compatibility check to all compatibility checks that were changed in a project:
\begin{equation*}
	\frac{\#added}{\#added+\#deleted}
\end{equation*}
We expected each project to have a value $\geq$0.5, since one cannot delete more lines than have been added.
However, for one project this value was below 0.5.
We found that this was caused by Git's branching mechanism:
The developers created two separate branches, removed the same compatibility checks in both, and merged them.
As a result, the merged history contains multiple commits deleting the same compatibility checks.
Thus, we cannot ensure that the actual number of compatibility checks is the balance of added and deleted checks.
Still, for all other projects, our assumption held true, supporting our confidence in the data.
Moreover, \dataset{} does not contain the respective guarded method calls, and thus additional analyses are required (e.g., control-flow analysis). 
In our replication package, we provide a script using a regular expression to extract the code section succeeding the compatibility check. Since it is a coarse method and to avoid copyright and licensing issues~\cite{Ballhausen2019, Baltes2019}, the extracted sections are not part of the dataset.

After this validation, we executed our script on all 2,399 Git repositories we crawled.
We inferred commits until the \texttt{author\_date} timestamp of January \nth{6}, 2021 (11:59:59 pm AoE).
Our script required roughly one day and we obtained 80,324 individually changed compatibility checks from 21,658 commits of 1,394 repositories.
We could extract Android versions and respective comparison operators for 78,503 ($\approx$97.7\,\%) of these compatibility checks.

\subsection{Dataset Description}

\begin{table}
	\caption{Fields of \dataset{}.}
	\label{tb:field-desc}
	\vspace*{-2ex}
	\centering \begin{tabular}{p{.3\linewidth}p{.6\linewidth}}
		\toprule
		\textbf{field} & \textbf{description}\\
		\midrule
		\texttt{repo\_name} & name of the repository\\
		\midrule[0pt]
		\texttt{repo\_url} & URI of the repository\\
		\midrule[0pt]
		\texttt{commit\_hash} & hash of the commit\\
		\midrule[0pt]
		\texttt{timestamp} & timestamp of the commit (i.e., the author date and time)\\
		\midrule[0pt]
		\texttt{old\_path} & the old path of the changed file (empty if the file was added in the respective commit)\\
		\midrule[0pt]
		\texttt{new\_path} & the new path of the changed file (empty if the file was deleted in the respective commit)\\
		\midrule[0pt]
		\texttt{change\_action} & indicator whether the compatibility check was added (i.e., \texttt{+++}) or deleted (i.e., \texttt{----})\\
		\midrule[0pt]
		\texttt{line} & the line containing the compatibility check\\
		\midrule[0pt]
		\texttt{version} & the extracted version code from the \texttt{andorid.os.Build.VERSION\_CODES}-package or \texttt{NONE\_DETECTED}\\
		\midrule[0pt]
		\texttt{compare\_type} & comparison used for the compatibility check (i.e.  \texttt{<|<=|==|>=|>} or \texttt{NONE\_DETECTED})\\
		\midrule[0pt]
		\texttt{timestamp\_ign\_tz} & same as \texttt{timestamp}, but drop timezone\\
		\bottomrule
	\end{tabular}
	\vspace*{-3.5ex}
\end{table}

\noindent
\dataset{} is a single csv file containing all changed compatibility checks as well as metadata of the commit and code. 
We summarize the individual fields in \autoref{tb:field-desc}.
For replication, we publish all of our artifacts.\textsuperscript{\ref{fn:repo}}

We denote whether the line containing the compatibility check was added (\texttt{+++}) or deleted (\texttt{----}) in the field \texttt{change\_action}.
If the compatibility check was changed (i.e., the code was modified), at least two entries (an addition and a deletion) for the same commit hash and source file (\texttt{new\_path}) exist.
Comparing \texttt{old\_path} and \texttt{new\_path} allows to determine whether an addition or deletion is a result of a new, removed, or renamed file.
The field \texttt{compare\_type} represents the normalized operator (i.e., \texttt{\{<|<=|==|>=|>\}}), whereas the field \texttt{version} is the extracted version code.
We further provide the \texttt{timestamp} of each commit based on the local time with and without timezone information (i.e., \texttt{timestamp\_ign\_tz}).
To not violate any developer's privacy, we deliberately left out any further information related to them (e.g., e-mail address, name, commit message).
As long as the repositories remain publicly available, researchers can analyze such and additional information using the repositories' Git commit hashes (\texttt{commit\_hash}).
\section{Conclusion}
\label{sec:usage}

\looseness=-1
\noindent
With \dataset{}, we aim to foster research on analyzing, detecting, and correcting Android incompatibilities.
Since \dataset{} comprises historical and evolutionary information, it allows researchers to investigate change patterns of compatibility checks.
As an example of such an analysis, we display a first overview of the evolution of the compatibility checks in \autoref{fig:changed-version}.
To simplify this overview, we merged the API levels into their respective versions, for instance, \textsc{ECLAIR\_0\_1} and \textsc{ECLAIR\_MR1} become \textsc{ECLAIR}.
We can see that certain versions, namely, \textsc{Honeycomb}, \textsc{Lollipop}, \textsc{Marshmallow} (M), and \textsc{Oreo} (O), are involved in more changed compatibility checks than other versions.
For example, even in 2020, compatibility checks for \textsc{Honeycomb} are changed, even though it is not maintained anymore since 2016.

Building on \dataset{}, we envision several directions for future research, such as:

\noindent\textbf{Pattern Detection.} Detecting incompatibilities is a helpful means to improve the quality of software and avoid design flaws.
Deriving patterns (i.e., similar to error patterns of Livshits and Zimmermann~\cite{Livshits2005}) can help to improve such detection in an automated fashion, not only in the current code base, but also throughout a system's history.
For instance, it has been shown that compatibility checks can help to detect device-induced~\cite{Wei2019} or callback-caused issues~\cite{Huang2018}.
Such research helps to understand incompatibilities, their evolution, and impact, as well as the design of novel techniques.

\noindent\textbf{Automated Program Repair.} In general, Android incompatibilities can be considered as a form of an API misuse.
We may be able to use and cluster identified patterns (i.e., similar compatibility checks) to automatically identify design flaws in the code (e.g., security issues due to unguarded API calls)~\cite{Zhong2009,Nielebock2020a}.
Moreover, we can use the historical information to identify past fixes of incompatibilities and reuse them to automatically repair the same incompatibilities in other code locations, similar to research on automated bug repair~\cite{Kim2006,Sun2010,Le2016,Long2016}.

\noindent\textbf{Benchmarking.} Finally, \dataset{} can serve as a dataset for benchmarking new tools, for instance, for automated program repair.
Novel tools may exploit different techniques that can be evaluated and compared based on \dataset{}.
As a concrete example, we are working on a technique for cooperative program repair~\cite{Nielebock2020}, for which we aim to use \dataset{} to evaluate its performance.

\noindent \textbf{Acknowledgments.} This research has been supported by the German Research Foundation (SA 465/49-3).

\balance
\bibliographystyle{IEEEtranS}
\bibliography{MYfull,paper-datashowcase-android-misuse}

\end{document}